\documentclass[11pt]{article}
\usepackage{epsfig} 
\newcommand{\Nf}{N_{\!f}} 
\newcommand{\MSbar}{\overline{\mbox{MS}}} 

\begin{document}
\title{Three loop $\MSbar$ tensor current anomalous dimension in QCD} 
\author{J.A. Gracey, \\ Theoretical Physics Division, \\ Department of 
Mathematical Sciences, \\ University of Liverpool, \\ Peach Street, \\ 
Liverpool, \\ L69 7ZF, \\ United Kingdom.} 
\date{} 
\maketitle 
\vspace{5cm} 
\noindent 
{\bf Abstract.} The anomalous dimensions of the general flavour non-singlet
quark bilinear currents $\bar{\psi} \gamma^{[\mu_1} \ldots \gamma^{\mu_n]} 
\psi$ are computed at three loops in quantum chromodynamics in the minimal 
subtraction scheme. The dimension of the tensor current emerges for the case
$n$ $=$ $2$ and the anomalous dimension for the general flavour singlet 
current is also discussed.  

\vspace{-16cm} 
\hspace{13.5cm} 
{\bf LTH 482} 

\newpage 
In recent years the construction of perturbative results at large numbers of
loops in quantum chromodynamics, (QCD), has advanced substantially. For
example, the $\beta$-function and various other renormalization group functions
are now known to four loops in the $\MSbar$ scheme, \cite{1,2,3}. One 
motivation for such calculations rests in essence with the improvement in
precision of experiments relating to the strong nuclear force. Whilst such 
four loop results represent the current limit of our ability to calculate in 
perturbative QCD, various quantities remain to be determined at three loops.
For instance, whilst the anomalous dimension of various quark bilinear currents
such as the axial vector current are known to at least three loops, 
\cite{4,5,6}, that of the closely related flavour non-singlet tensor current  
$\bar{\psi} \sigma^{\mu\nu} \psi$ where $\sigma^{\mu\nu}$~$=$~$[\gamma^\mu,
\gamma^\nu]$ and $\psi$ is the quark field, is only known at two loops. This 
anomalous dimension occurs in the matching between currents in QCD heavy quark 
effective theory, \cite{7}. In \cite{7}, the two loop $\MSbar$ result was 
deduced indirectly as opposed to performing an explicit two loop 
renormalization. In this letter we complete the gap in the literature by 
providing the value of the three loop $\MSbar$ anomalous dimension of the 
current $\bar{\psi} \sigma^{\mu\nu} \psi$. While this may appear to represent a
moderate progression in this area, we choose to determine it not by explicit 
calculation of the current on its own but deduce it as a corollary of the 
renormalization of a set of generalized currents. These are 
$j_{(n)}^{\mu_1 \ldots \mu_n}$ $=$ $\bar{\psi} \Gamma_{(n)}^{\mu_1 \ldots 
\mu_n} \psi$, where $\Gamma_{(n)}^{\mu_1 \ldots \mu_n}$ is the antisymmetric 
product of $n$ $\gamma$-matrices. Clearly for QCD only the first five currents 
are non-zero but our motivation for considering the larger set relates to other
issues. First, it is clear that for $n$ $=$ $2$ the result we seek for the 
tensor current will emerge simply. However, by calculating with the general 
currents the results for $n$ $=$ $0$ and $1$ and the naive results for $n$ $=$ 
$3$ and $4$ will provide non-trivial checks on the calculation from the point 
of view of, say, symmetry factors and correctly including all the three loop 
Feynman diagrams. These three loop results have been determined in 
\cite{8,9,5}. We have refered to the results for $n$ $=$ $3$ and $4$ as naive 
since in these cases there is a connection with the $\gamma^5$ problem of 
dimensional regularization. In particular the anomalous dimensions which emerge
for $n$ $=$ $3$ and $4$ do not correspond directly with the anomalous 
dimensions of the axial vector and pseudoscalar currents respectively. The 
correct anomalous dimensions for these currents are determined by including a 
finite renormalization to ensure that chiral symmetry is preserved, \cite{5}. 
Such a procedure was elaborated on in detail in \cite{5} but this issue in the 
context of the matrices $\Gamma_{(n)}$ needs to be addressed due to its 
potential application to other operators such as those which contain four quark
fields and one or more $\gamma^5$-matrices. One reason for this is that the 
algebra of the $\Gamma_{(n)}$-matrices has been widely studied, 
\cite{10,11,12,13}, and also provides a simpler way to programme in a symbolic 
manipulation language than say the split $\gamma^5$ algebra of \cite{14}. 
Indeed the renormalization of $j_{(n)}$ at three loops which we carry out here 
is performed with the use of the {\sc Mincer} package, \cite{15}, written in 
the language {\sc Form}, \cite{16}. While the above motivations for this work 
have been in relation to QCD, we note that knowledge of the anomalous dimension
of the currents $j_{(n)}$ are necessary for other problems. For instance, the 
relation between QCD and the non-abelian Thirring model, (NATM), in 
$d$-dimensions has been developed in \cite{17} based on the earlier 
observations of \cite{18}. To understand the connection further, one needs not 
only to have knowledge of the fundamental renormalization group functions such 
as the $\beta$-function, but also the anomalous dimensions of the operators in 
both theories as well. By providing these for the $n$-dependent operators 
additional information is being determined for this area. For example, in the 
strictly two dimensional NATM the currents with $n$ $=$ $3$ and $4$ would be 
zero. However, in the $d$-dimensional context of the fixed point equivalence 
with QCD, \cite{17,18}, they would be evanescent and important for the 
connection since from the QCD point of view they do not vanish in the limit to 
four dimensions. 

We now turn to the discussion of the computation. The procedure to renormalize
the currents $j_{(n)}^{\mu_1 \ldots \mu_n}$ is to insert the operator in the
quark two point function, use dimensional regularization and determine the 
poles in $\epsilon$, where $d$ $=$ $4$ $-$ $2\epsilon$, before minimally 
subtracting them. In particular the set of Feynman diagrams to three loops are
generated with the {\sc Qgraf} package, \cite{19}, and converted into {\sc 
Form} input for processing with the {\sc Mincer} procedures, \cite{15}. As this
is a well documented method we merely note the major issues in relation to the 
operators we are interested in. First, by constructing currents with the 
generalized $\gamma$-matrices $\Gamma_{(n)}^{\mu_1 \ldots \mu_n}$ defined by 
\begin{equation} 
\Gamma_{(n)}^{\mu_1 \ldots \mu_n} ~=~ \gamma^{[\mu_1} \ldots \gamma^{\mu_n]}
\end{equation} 
there is no mixing under renormalization between any of the $j_{(n)}^{\mu_1 
\ldots \mu_n}$. Next to apply the {\sc Mincer} routines the Green's function 
must be converted to scalar integrals. This requires multiplying the quark two 
point function by $\Gamma_{(n) \, \mu_1 \ldots \mu_n}$, taking the spinor trace
and dividing by the normalization $\mbox{tr} (\Gamma_{(n)}^{\mu_1 \ldots \mu_n}
\Gamma_{(n) \, \mu_1 \ldots \mu_n})$. As we are only interested in flavour 
non-singlet currents for the moment, the $\gamma$-matrix strings contain either
two $\Gamma_{(n)}$'s or none. For those with none the trace operations is 
readily performed. For the former strings one has to use the properties of the 
$\Gamma_{(n)}^{\mu_1 \ldots \mu_n}$ which are given in \cite{11,12,13}. It 
transpires that we only need a subset of those results. By considering the 
Feynman diagrams which arise at three loops it is clear that the two 
$\Gamma_{(n)}$'s are separated by at most six ordinary $\gamma$-matrices. 
Therefore, from the general expression, \cite{12},  
\begin{equation} 
\Gamma^{\mu_1 \ldots \mu_n}_{(n)} \Gamma^{\nu_1 \ldots \nu_m}_{(m)}  
\Gamma_{(n)\, \mu_1 \ldots \mu_n} ~=~ f(n,m) \Gamma^{\nu_1 \ldots \nu_m}_{(m)}  
\end{equation}  
it is easy to deduce 
\begin{equation} 
\Gamma^{\mu_1 \ldots \mu_n}_{(n)} \gamma^\mu \gamma^\nu 
\Gamma_{(n)\, \mu_1 \ldots \mu_n} ~=~ f(n,2) \gamma^\mu \gamma^\nu ~+~ 
(f(n,0) - f(n,2)) \eta^{\mu\nu} 
\end{equation}  
and 
\begin{eqnarray} 
\Gamma^{\mu_1 \ldots \mu_n}_{(n)} \gamma^\mu \gamma^\nu \gamma^\sigma 
\gamma^\rho \Gamma_{(n)\, \mu_1 \ldots \mu_n} &=& f(n,4) \gamma^\mu \gamma^\nu
\gamma^\sigma \gamma^\rho \nonumber \\ 
&& +~ (f(n,2) - f(n,4)) \left[ \eta^{\mu\nu} \gamma^\sigma \gamma^\rho 
- \eta^{\mu\sigma} \gamma^\nu \gamma^\rho + \eta^{\mu\rho} \gamma^\nu 
\gamma^\sigma \right. \nonumber \\
&& \left. ~~~~~~~~~~~~~~~~~~~~~~~~~~~~ + \eta^{\sigma\rho} \gamma^\mu 
\gamma^\nu - \eta^{\nu\rho} \gamma^\mu \gamma^\sigma + \eta^{\nu\sigma} 
\gamma^\mu \gamma^\rho \right] \nonumber \\
&& +~ (f(n,4) - 2f(n,2) + f(n,0)) \left[ \eta^{\mu\nu} \eta^{\sigma\rho} 
- \eta^{\mu\sigma} \eta^{\nu\rho} + \eta^{\mu\rho} \eta^{\nu\sigma} \right] ~. 
\nonumber \\ 
\end{eqnarray}  
The analogous relation involving six $\gamma$-matrices between two contracted 
$\Gamma_{(n)}$'s is too large to quote here though it clearly will involve the 
function $f(n,6)$. The values of $f(n,m)$ we require are deduced from the 
general formula given in \cite{12}. We find 
\begin{eqnarray} 
f(n,2) &=& \left[ d^2 - d - 4dn + 4n^2 \right] \frac{f(n,0)}{d(d-1)}  
\nonumber \\ 
f(n,4) &=& \left[ d^4 - 8d^3n - 6d^3 + 24d^2n^2 + 24d^2n + 11d^2 - 32dn^3 
- 24dn^2 \right. \nonumber \\ 
&& \left. ~ - 32dn - 6d + 16n^4 + 32n^2 \right] 
\frac{f(n,0)}{d(d-1)(d-2)(d-3)}  \nonumber \\ 
f(n,6) &=& \left[ 1 ~-~ 12 \frac{(d-n)}{d} ~+~ 60 \frac{(d-n)(d-n-1)}{d(d-1)} 
{}~-~ 160 \frac{(d-n)(d-n-1)(d-n-2)}{d(d-1)(d-2)} \right. \nonumber \\
&& \left. ~+~ 240 \frac{(d-n)(d-n-1)(d-n-2)(d-n-3)}{d(d-1)(d-2)(d-3)} \right.
\nonumber \\
&& \left. ~-~ 192 \frac{(d-n)(d-n-1)(d-n-2)(d-n-3)(d-n-4)} 
{d(d-1)(d-2)(d-3)(d-4)} \right. \nonumber \\ 
&& \left. ~~+~ 64 \frac{(d-n)(d-n-1)(d-n-2)(d-n-3)(d-n-4)(d-n-5)} 
{d(d-1)(d-2)(d-3)(d-4)(d-5)} \right] f(n,0) ~. \nonumber \\   
\end{eqnarray} 
They have been written in terms of $f(n,0)$ since this also occurs in the 
normalizing factor and hence will cancel in the computation. With these lemmas
the $\Gamma_{(n)}$'s can be removed from all $\gamma$-strings and the spinor
trace evaluated in the normal fashion which then means the integrals are in the
correct format for applying the {\sc Mincer} procedures. 

Taking into account the three loop wave function renormalization, \cite{20}, of
the quark fields present in the original operator, we find the following result
for the anomalous dimension of $j_{(n)}$ for arbitrary $n$,  
\begin{eqnarray} 
\gamma_{(n)}(a) &=& -~ (n-1)(n-3) C_F a \nonumber \\
&& +~ \left[ 4(n-15)T_F\Nf ~+~ (18n^3 - 126n^2 + 163n + 291 )C_A \right. 
\nonumber \\ 
&& \left. ~~~~~-~ 9(n-3)( 5n^2 - 20n + 1)C_F \right] 
\frac{(n-1) C_F a^2}{18}  \nonumber \\ 
&& +~ \left[ \left( 144n^5 - 1584n^4 + 6810n^3 - 15846n^2 + 15933n + 11413 
\right. \right. \nonumber \\ 
&& \left. \left. ~~~~~ - 216n(n-3)(n-4)(2n^2-8n+13)\zeta(3) \right) C_A^2 
\right. \nonumber \\
&& \left. ~~~~~-~ \left( 3(72n^5 - 792n^4 + 3809n^3 - 11279n^2 + 15337n 
+ 1161 ) \right. \right. \nonumber \\ 
&& \left. \left. ~~~~~~~~~~~-~ 432n(n-3)(n-4)(3n^2-12n+19)\zeta(3) \right) 
C_A C_F \right. \nonumber \\
&& \left. ~~~~~+~ \left( 8(3n^3 + 51n^2 - 226n - 278 ) ~+~ 1728(n-3)\zeta(3) 
\right) C_AT_F\Nf \right. \nonumber \\ 
&& \left. ~~~~~-~ \left( 18(n-3)(17n^4 - 136n^3 + 281n^2 - 36n + 129) \right. 
\right. \nonumber \\ 
&& \left. \left. ~~~~~~~~~~~+~ 864n(n-3)(n-4)(n^2 - 4n + 6)\zeta(3) \right) 
C_F^2 \right. \nonumber \\ 
&& \left. ~~~~~-~ \left( 12(17n^3 + n^2 - 326n + 414) ~+~ 1728(n-3)\zeta(3) 
\right) C_F T_F\Nf \right. \nonumber \\ 
&& \left. ~~~~~+~ 16(13n - 35) T_F^2 \Nf^2 \right] 
\frac{(n-1) C_F a^3}{108} ~+~ O(a^4) ~.  
\label{anomdim} 
\end{eqnarray}  
where $a$ $=$ $\alpha_s/4\pi$ $=$ $g^2/(16\pi^2)$ is the coupling constant, 
$T_F$, $C_F$ and $C_A$ are the usual colour group Casimirs and $\Nf$ is the 
number of quark flavours. There are various checks on this result. First, as 
the operator itself is gauge invariant, its anomalous dimension must be 
independent of the covariant gauge parameter, $\xi$. Therefore, in the 
calculation we used a gluon propagator of the form $(\eta_{\mu\nu} - \xi 
p_\mu p_\nu/p^2)/p^2$ where $p$ is the momentum, and observed that $\xi$ 
cancelled in the final result, (\ref{anomdim}). Second, the known three loop
$\MSbar$ results for $n$ $=$ $0$, \cite{8,9}, and $1$ emerge. In the latter 
case the anomalous dimension vanishes as the current corresponds to the 
conserved electric current. Moreover, the value of (\ref{anomdim}) for $n$ $=$ 
$3$ and $4$ corresponds to the naive values obtained by Larin in \cite{5}. 
Hence, we can deduce that the anomalous dimension for the tensor current 
$\bar{\psi} \sigma^{\mu\nu} \psi$ is  
\begin{eqnarray} 
\gamma_{(2)}(a) &=& C_F a ~+~ [ 257C_A ~-~ 171C_F ~-~ 52 T_F \Nf ] 
\frac{C_F a^2}{18} \nonumber \\ 
&& +~ \left[ 13639 C_A^2 ~-~ 4320\zeta(3) C_A^2 ~+~ 
12096\zeta(3)C_A C_F \right. \nonumber \\ 
&& \left. ~~~~-~ 20469 C_A C_F ~-~ 1728\zeta(3) C_A T_F \Nf ~-~ 
4016 C_A T_F \Nf \right. \nonumber \\ 
&& \left. ~~~~-~ 6912 \zeta(3) C_F^2 ~+~ 6570 C_F^2 ~+~ 
1728\zeta(3) C_F T_F \Nf \right. \nonumber \\ 
&& \left. ~~~~+~ 1176 C_F T_F \Nf ~-~ 144T_F^2 C_F^2) \right] 
\frac{C_F a^3}{108} ~+~ O(a^4)  
\label{anomdimtens} 
\end{eqnarray}  
by substituting $n$ $=$ $2$ in (\ref{anomdim}). The first two terms are in 
agreement with \cite{7}. When the gauge group is $SU(3)$ we have 
\begin{eqnarray} 
\gamma_{(2)}(a) &=& \frac{4a}{3} ~-~ \frac{2[26\Nf - 543]a^2}{27} \nonumber \\
&& -~ \left[ 36\Nf^2 + 1440\zeta(3) \Nf + 5240\Nf + 2784\zeta(3) - 52555 
\right] \frac{a^3}{81} ~+~ O(a^4)  
\end{eqnarray}  
or numerically 
\begin{eqnarray} 
\gamma_2(a) &=& 1.333333a ~-~ ( 1.925926\Nf - 40.222222 )a^2 \nonumber \\
&& -~ ( 0.444444\Nf^2 + 86.061259\Nf - 607.512020 ) a^3 ~+~ O(a^4) ~.  
\end{eqnarray}  
We note that whilst we have concentrated on the flavour non-singlet current,
the expression (\ref{anomdimtens}) also corresponds to the anomalous dimension
of the (anomaly free) flavour singlet tensor current. This follows trivially
since the Feynman diagrams with a closed quark loop with one $\bar{\psi} 
\sigma^{\mu\nu} \psi$ operator insertion in it, has an odd number of 
$\gamma$-matrices and are therefore zero upon taking the spinor trace.  

One question which naturally arises out of choosing to compute these anomalous
dimensions with the generalized $\gamma$-matrices, $\Gamma_{(n)}$, is that of
how they relate to currents which involve $\gamma^5$. As is well known its 
treatment in dimensional regularization is a technically difficult exercise due
to the fact that it has no natural $d$-dimensional generalization, \cite{14}. 
In \cite{5} this issue of renormalizing the axial vector and pseudoscalar 
currents within the automatic multiloop computer algebra approach was 
addressed. To treat such currents one first of all defines $\gamma^5$ in terms 
of $\epsilon_{\mu\nu\sigma\rho} \Gamma_{(4)}^{\mu\nu\sigma\rho}$ where 
$\epsilon_{\mu\nu\sigma\rho}$ is the totally antisymmetric four dimensional 
pseudotensor, \cite{5}. Then products of the $\epsilon$-tensor are replaced by 
a function of the metric, $\eta_{\mu\nu}$, which does have a natural 
$d$-dimensional extension. To compensate for the lack of continuity of the 
$\gamma^5$ definition in $d$-dimensions a finite renormalization is performed 
in addition to the usual $\MSbar$ subtraction of the Green's function, 
\cite{5}. The condition for the finite renormalization is to impose the obvious
anticommutativity of $\gamma^5$ on the finite renormalized Green's functions. 
We have summarized the renormalization of these quark currents of previous 
work, \cite{5}, to allow the interested reader to contrast their 
renormalization here. Clearly within the context of the complete set of 
currents $j_{(n)}^{\mu_1 \ldots \mu_n}$ the correct anomalous dimension for the
axial vector and pseudoscalar currents should somehow be present in our 
results. We now address this. 

In choosing to work in a spacetime which is $d$-dimensional, with $d$ 
non-integer, the problem of defining $\gamma^5$ is in some sense bypassed in 
that one can regard it as an object which does not exist naturally. Also the 
ordinary $\gamma$-matrix basis ceases to be {\em finite} dimensional. Indeed in
this situation the $\Gamma_{(n)}$-matrices provide a more natural basis for 
performing $d$-dimensional calculations. (For a more detailed discussion on 
this point see, for example, \cite{12,13,21}.) As $\gamma^5$ is a manifestly 
four dimensional object one need only be concerned about incorporating its 
effect when projecting from the $d$-dimensional spacetime, with its infinite
$\Gamma_{(n)}$-basis, onto the integer dimensional physical spacetime. For us 
this projection is not achieved by analytically continuing the product 
$\epsilon_{\mu\nu\sigma\rho} \epsilon^{\alpha\beta\delta\lambda}$ to 
$d$-dimensional spacetime, \cite{5}. Instead within the context of the infinite
dimensional $\gamma$-matrix basis one determines a finite renormalization 
constant $Z_{\mbox{\footnotesize{fin}}}$ from a condition similar to that of 
\cite{5}, 
\begin{equation} 
G^{(2)}(p^2,n) ~=~ Z_{\mbox{\footnotesize{fin}}} G^{(2)}(p^2,4-n) 
\label{zfindef} 
\end{equation} 
where $G^{(2)}(p^2,n)$ is the finite part of the quark two point Green's 
function after minimal subtraction where the current $j_{(n)}^{\mu_1 \ldots
\mu_n}$ has been inserted. In other words it is the renormalized Green's 
function. Also in (\ref{zfindef}) it is understood that all the contributing 
diagrams have been multiplied by $\Gamma_{(n) \, \mu_1 \ldots \mu_n}$ before 
taking the spinor trace and dividing by the appropriate normalization mentioned
earlier. Choosing the argument of the right side of (\ref{zfindef}) to be 
$(4-n)$ ensures that for instance the finite renormalization constant which 
emerges for the $n$ $=$ $4$ current will give the same anomalous dimension as 
the $n$ $=$ $0$ current with no finite renormalization. Therefore from our 
calculations we find that
\begin{eqnarray} 
Z_{\mbox{\footnotesize{fin}}} &=& 1 ~+~ 4(n - 2) C_F a \nonumber \\
&& -~ \left[ (36n^2 - 144n + 1) C_A 
- 18n(5n - 16)C_F + 4 T_F \Nf \right] \frac{(n - 2) C_F a^2}{9} \nonumber \\ 
&& +~ \left[ 216(6n^4 - 48n^3 + 134n^2 - 152n + 39) \zeta(3) C_A^2 \right. 
\nonumber \\
&& \left. ~~~~~-~ 3(144n^4 - 1152n^3 + 3292n^2 - 3952n - 479) C_A^2 \right. 
\nonumber \\  
&& \left. ~~~~~+~ 216(6n^4 - 48n^3 + 134n^2 - 152n + 39) \zeta(3) C_A^2 \right. 
\nonumber \\ 
&& \left. ~~~~~+~ 6(108n^4 - 1080n^3 + 4169n^2 - 6314n + 200) C_A C_F \right. 
\nonumber \\  
&& \left. ~~~~~-~ 432 (9n^4 - 72n^3 + 200n^2 - 224n + 57) \zeta(3) C_A C_F 
\right. \nonumber \\ 
&& \left. ~~~~~-~ 8(6n^2 - 24n + 107) C_A T_F \Nf ~-~ 
1728 \zeta(3) C_A T_F \Nf \right. \nonumber \\ 
&& \left. ~~~~~+~ 54(17n^4 - 76n^3 - 32n^2 + 352n - 76) C_F^2 \right. 
\nonumber \\ 
&& \left. ~~~~~+~ 2592(n^4 - 8n^3 + 22n^2 - 24n + 6) \zeta(3) C_F^2 \right. 
\nonumber \\  
&& \left. ~~~~~+~ 12(34n^2 - 148n + 145) C_F T_F \Nf \right. \nonumber \\
&& \left. ~~~~~+~ 1728 \zeta(3) C_F T_F \Nf ~-~ 208 T_F^2 \Nf^2 \right] 
\frac{(n-2) C_F a^3}{81} ~+~ O(a^4)  
\label{zfin} 
\end{eqnarray} 
and we note that it is independent of the gauge fixing parameter. Moreover, by 
construction it evaluates to unity for $n$ $=$ $2$ as it ought. Hence, to 
deduce the anomalous dimension of the currents $\bar{\psi} \gamma^\mu \gamma^5 
\psi$ and $\bar{\psi} \gamma^5 \psi$ themselves by this method, one must
therefore add the piece 
\begin{eqnarray} 
\mu \frac{d \ln Z_{\mbox{\footnotesize{fin}}} }{d \mu} &=&  
-~ \frac{4 C_F (11 C_A - 4 T_F \Nf )(n-2)a^2}{3} \nonumber \\ 
&& +~ 2C_F \left[ (396n^2 - 1584n - 601)C_A^2 ~-~ 
198(5n^2 - 20n + 8)C_F C_A \right. \nonumber \\ 
&& \left. ~~~~~~~~~~-~ 16 T_F^2 \Nf^2 ~-~ 16(9n^2 - 36n - 25) C_A T_F \Nf 
\right. \nonumber \\
&& \left. ~~~~~~~~~~+~ 72(5n^2 - 20n + 11) C_F T_F \Nf \right] 
\frac{(n-2)a^3}{27} ~+~ O(a^4) 
\label{finren} 
\end{eqnarray} 
to $\gamma_{(3)}(a)$ and $\gamma_{(4)}(a)$ respectively for $n$ $=$ $3$ and $4$
where $\mu$ is the renormalization scale. This effectively projects out the 
true component of (\ref{anomdim}) for four dimensional spacetime and correctly
reproduces the known results of \cite{4,5}. We note that the three loop 
contribution of (\ref{finren}) only involves the two loop term of (\ref{zfin}).
Also if one was working with reference to two dimensions the argument of the 
right side of (\ref{zfindef}) would instead be $(2-n)$. 

Finally, we briefy comment on this approach to flavour singlet current 
anomalous dimensions which we have also analyzed. As noted earlier the flavour
singlet and non-singlet currents coincide for even $n$. However, if one were to
use the $\Gamma_{(n)}$-basis to study the anomalous dimension of singlet 
currents for $n$ odd then there are two further issues to be dealt with. The 
first is the evaluation of $\gamma$-strings with only one $\Gamma_{(n)}$-matrix
in the spinor trace. To handle these diagrams the following results are 
necessary. First, one decomposes the $\gamma$-string arising from the 
propagator and vertices into the $\Gamma_{(n)}$-basis and then uses the general
property that
$\mbox{tr} (\Gamma^{\mu_1 \ldots \mu_n}_{(n)} \Gamma_{(p)}^{\nu_1 \ldots 
\nu_p}) \mbox{tr} (\Gamma_{(n) \, \mu_1 \ldots \mu_n} 
\Gamma_{(q)}^{\sigma_1 \ldots \sigma_q})$ is proportional to $\delta_{pq}$. The
tensor of proportionality involves a linear combination of products of 
$(p+q)/2$ $\eta$-tensors which respect the antisymmetry of the Lorentz indices
of $\Gamma_{(p)}$ and $\Gamma_{(q)}$. At three loops the largest value of $p$
or $q$ is $5$ which means that a small set of traces needs to be deduced. This
is achieved by noting that, \cite{12,13},  
\begin{equation}  
\mbox{tr} \left(\Gamma^{\mu_1 \ldots \mu_m}_{(m)} \Gamma_{(n)}^{\nu_1 \ldots 
\nu_n} \right) ~=~ 4 (-1)^{n(n-1)/2} \delta_{mn} n ! \left[ \eta^{\mu_1\nu_1} 
\ldots \eta^{\mu_n\nu_n} ~+~ \mbox{antisymmetric permutations} \right] ~. 
\end{equation}
With these identities the graphs with a single operator insertion can be 
determined and the basic anomalous dimension analogous to (\ref{anomdim}) 
deduced as 
\begin{eqnarray} 
\gamma^{\mbox{\footnotesize{singlet}}}_{(n)}(a) &=& \gamma_{(n)}(a) ~+~ 
12 \delta_{n,3} C_F T_F \Nf a^2 \nonumber \\ 
&& +~ \left[ \delta_{n,3} C_F \left( \frac{218}{3} C_A T_F \Nf  ~+~ 
\frac{8}{3} T_F^2 \Nf^2 -~ 36 C_F T_F \Nf \right) \right. 
\nonumber \\ 
&& \left. ~~~~~+~ \delta_{n,5} (480 \zeta(3) + 80) 
\frac{d_F^{abc} d_F^{abc}}{N_{\mbox{\footnotesize{fund}}} } \right] a^3 ~+~ 
O(a^4)  
\end{eqnarray} 
where 
\begin{equation} 
d_F^{abc} ~=~ \mbox{Tr} \left( T^{(a} T^b T^{c)} \right) ~, 
\end{equation} 
$T^a$ are the colour group generators and $N_{\mbox{\footnotesize{fund}}}$ is
the dimension of its fundamental representation. The $\delta$-symbols arise 
from the spinor traces with one $\Gamma_{(n)}$-matrix. Whilst an additional 
colour group Casimir enters for the case $n$ $=$ $5$, this current is 
evanescent with respect to four dimensions and is therefore not physically 
important. Clearly to relate the $n$ $=$ $3$ current to the four dimensional 
axial vector current anomalous dimension one requires an additional finite 
renormalization and this arises from two places. The first occurs by imposing 
the condition (\ref{zfindef}) and deserves little more comment only to record 
that an extra term arises in (\ref{finren}) when $n$ $=$ $3$ which is 
\begin{equation} 
2 C_F \left[ 77 C_A T_F \Nf ~-~ 28 T_F^2 \Nf^2 \right] a^3 ~. 
\end{equation} 
However, this additional finite renormalization will not ensure the correct
singlet anomalous dimension emerges since the (four dimensional) chiral anomaly
will not be preserved. This requires another finite renormalization and is 
achieved here by the method developed in \cite{5} but with the anomaly equation
treated inside a quark two point function in contrast to the gluon two point 
function considered in \cite{5}. As this procedure only involves the gluonic 
operator inserted in a one loop diagram to the order we are interested in, we 
note that the additional contribution will be  
\begin{equation} 
2 C_F \left[ -~ 66 C_A T_F \Nf ~+~ 24 T_F^2 \Nf^2 \right] a^3 
\end{equation} 
which gives the correct singlet axial vector current anomalous dimension, 
\cite{4,5}. 
 
In conclusion we have provided a new term in the series for the anomalous 
dimension of the tensor current in QCD in the $\MSbar$ scheme. Also by 
considering the generalized $\gamma$-matrix basis, $\Gamma_{(n)}$, we have 
demonstrated how the anomalous dimensions of the currents which involve 
$\gamma^5$ emerge. Indeed we believe this is an important aspect of the 
calculation as it in principle provides a more systematic and alternative way 
of renormalizing currents or other composite operators in QCD which involve the
purely four dimensional object $\gamma^5$. For example, one need only calculate
Green's functions with a general insertion and then the anomalous dimensions 
for a variety of operators will emerge by choosing the appropriate value of the
parameter of the inserted operator. It would be interesting, for instance, to 
develop this approach for more complicated and physically important operators 
such as the four quark operators which are fundamental to evaluating QCD 
corrections to weak processes. 

\vspace{1cm} 
\noindent 
{\bf Acknowledgements.} This work was supported in part by PPARC through an 
Advanced Fellowship. The author thanks Dr Austin G.M. Pickering for useful 
discussions. The calculations were performed with the use of the symbolic 
manipulation package {\sc Form}, \cite{16}.


\newpage 
\begin{thebibliography}{99} 
\bibitem{1} T. van Ritbergen, J.A.M. Vermaseren \& S.A. Larin, Phys. Lett. 
{\bf B400} (1997), 379. 
\bibitem{2} K.G. Chetyrkin, Phys. Lett. {\bf B404} (1997), 161; J.A.M. 
Vermaseren, S.A. Larin \& T. van Ritbergen, Phys. Lett. {\bf B405} (1997), 327. 
\bibitem{3} K.G. Chetyrkin \& A. R\'{e}tey, hep-ph/9910332.  
\bibitem{4} J. Kodaira, Nucl. Phys. {\bf B165} (1980), 129. 
\bibitem{5} S.A. Larin, Phys. Lett. {\bf B303} (1993), 113.  
\bibitem{6} S.A. Larin, T. van Ritbergen \& J.A.M. Vermaseren, Phys. Lett. {\bf
B404} (1997), 153.   
\bibitem{7} D.J. Broadhurst \& A.G. Grozin, Phys. Rev. {\bf D52} 
(1995), 4082.
\bibitem{8} D.V. Nanopoulos \& D.A. Ross, Nucl. Phys. {\bf B157} (1979), 273. 
\bibitem{9} O. Tarasov, JINR preprint P2-82-900.  
\bibitem{10} A.D. Kennedy, J. Math. Phys. {\bf 22} (1981), 1330.  
\bibitem{11} A. Bondi, G. Curci, G. Paffuti \& P. Rossi, Ann. Phys. {\bf 199} 
(1990), 268. 
\bibitem{12} A.N. Vasil'ev, S.\'{E}. Derkachov \& N.A. Kivel, Theor. Math. 
Phys. {\bf 103} (1995), 179. 
\bibitem{13} A.N. Vasil'ev, M.I. Vyazovskii, S.\'{E}. Derkachov \& N.A. 
Kivel, Theor. Math. Phys. {\bf 107} (1996), 27. 
\bibitem{14} G. 't Hooft \& M. Veltman, Nucl. Phys. {\bf B44} (1972), 189. 
\bibitem{15} S.G. Gorishny, S.A. Larin, L.R. Surguladze \& F.V. Tkachov,  
Comput. Phys. Commun. {\bf 55} (1989), 381. 
\bibitem{16} J.A.M. Vermaseren, ``{\sc FORM}'' version $2.3$, (CAN Amsterdam, 
1992). 
\bibitem{17} M. Ciuchini, S.\'{E}. Derkachov, J.A. Gracey \& A.N. Manashov, 
Nucl. Phys. {\bf B579} (2000), 56.  
\bibitem{18} A. Hasenfratz \& P. Hasenfratz, Phys. Lett. {\bf B297} (1992), 
166.
\bibitem{19} P. Nogueira, J. Comput. Phys. {\bf 105} (1993), 279.  
\bibitem{20} S.A. Larin \& J.A.M. Vermaseren, Phys. Lett. {\bf B303} (1993),
334.
\bibitem{21} J.C. Collins, {\it Renormalization} (Cambridge University Press,
1984). 
\end{thebibliography}
\end{document}